\documentclass[twocolumn,showpacs,preprintnumbers,amsmath,amssymb]{revtex4}

\usepackage{graphicx}
\usepackage{dcolumn}
\usepackage{bm}

\begin{document}


\title{Calibrating an interferometric laser frequency stabilization to MHz precision}

\author{Johannes F. S. Brachmann}
\affiliation{Max-Planck-Institut f\"ur Quantenoptik, 85748 Garching, Germany}
\altaffiliation[Also at ]{Ludwig-Maximilians-Universit\"at M\"unchen, 80799 M\"unchen, Germany}
\altaffiliation[and ]{Centre for Quantum Technologies, National University of Singapore, Singapore 117543}
\email{Hannes.Brachmann@mpq.mpg.de}

\author{Thomas Kinder}
 \email{Thomas.Kinder@tem-messtechnik.de}
\affiliation{TEM Messtechnik GmbH, 30559 Hannover, Germany}

\author{Kai Dieckmann}
\affiliation{Centre for Quantum Technologies, National University of Singapore, Singapore 117543}

\date{\today}

\begin{abstract}
We report on a calibration procedure that enhances the precision of an interferometer based frequency stabilization by several orders of magnitude. For this purpose the frequency deviations of the stabilization are measured precisely by means of a frequency comb. This allows to implement several calibration steps that compensate different systematic errors. The resulting frequency deviation is shown to be less than $5.7\,$MHz (rms $1.6\,$MHz) in the whole wavelength interval $750 - 795\,$nm. Wide tuning of a stabilized laser at this exceptional precision is demonstrated.
\end{abstract}

\maketitle

\section{Introduction}
Fabry-P\'erot interferometers have advanced to very high precision wavelength sensors \cite{high_finesse,scholl:3318}. In a Fizeau type setup and combined with an absolute frequency reference (typically an Helium-Neon laser) they are used as precision optical wavelength meters. By means of an additional feedback, stabilization and tuning of the laser wavelength can be accomplished \cite{PhysRevA.83.052515}. However, due to the comparatively long signal processing times, the feedback bandwidth of these systems is low and limiting the use for laser frequency stabilization.
\par
In this paper we present the optimization and calibration of an existing Fizeau type quadrature interferometer \cite{iScan_Patents} that is used for laser frequency stabilization and allows a very wide tuning range of a stabilized laser. Several orders of magnitude in precision over a wide wavelength range of about $45\,$nm are gained in comparison to an uncalibrated setup. The achieved accuracy with an rms deviation of $1.6\,$MHz is state-of-the-art interferometric laser frequency stabilization at high feedback bandwidth. In order to characterize and compensate the main observed frequency deviations occurring in the wavelength range $750\,-\,795\,$nm we use a fiber laser based frequency comb. We take these frequency deviations into account by a correction of the wavelength dependence of the refractive index up to second polynomial order. This allows for future calibration of the setup by employing only three interpolation points. For these measurements a Doppler-free saturation spectroscopy is easily used as an absolute frequency reference.
\begin{figure}[b!]
\centerline{\includegraphics[width=\columnwidth]{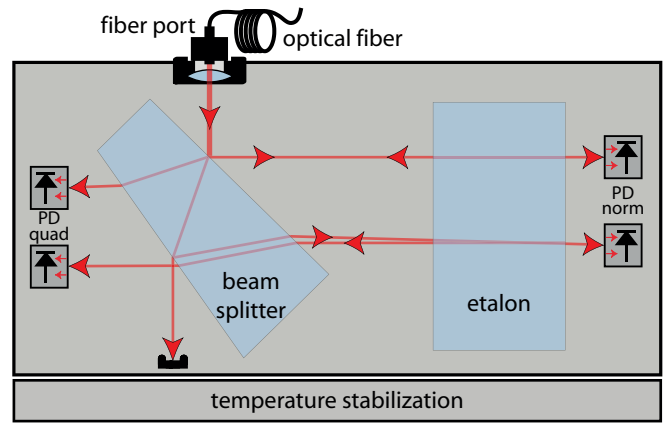}}
\caption{Patented interferometer setup subject to calibration. The interferometer is realized as an etalon of BK7 with a geometrical thickness of $50\,$mm. PD: photo detector, quad: quadrature, norm: normalization. See \cite{iScan_Patents} for details.}
\label{iscan_device}
\end{figure}
\section{Functional principle}
Prior to the description of the calibration of the interferometer, we only briefly review it's functional principle, as it is explained in detail elsewhere \cite{iScan_Patents}. Fig.\,\ref{iscan_device} shows a sketch of the interferometer setup. A test laser beam enters the device through an optical fiber and is split into two by means of a wedged beam splitter. The beams pass an etalon under a slight relative angle. Two pairs of photo diodes, one pair for each beam, are used to electronically construct two normalized periodic interferometer signals, $U_x$ and $U_y$, from the beams transmitted and reflected by the etalon \cite{iScan_Patents,BIRCH:90,HEYDEMANN:81}. The angle between the beams results in a $\pi/2$ relative phase shift between the interferometer signals. Due to the etalon's low finesse, their frequency dependence can be approximated by sine and cosine functions:
\begin{equation}
\label{phi_dependence}
\left(\begin{array}{c}
U_x\\
U_y\\
\end{array}\right)
\approx
\left(\begin{array}{c}
\mathrm{cos}\,\varphi \\
\mathrm{sin}\,\varphi \\
\end{array}\right)
\hspace{0.6cm} \textrm{with} \hspace{0.25cm} \varphi = 4 \pi \nu \frac{nL}{c_0}
\end{equation}
Hence, $U_x$ and $U_y$ can be used as quadrature signals describing a circular path in the x-y plane. Here, the interferometer phase $\varphi$ depends on the laser frequency $\nu$, and the thickness $L$ and the refractive index $n$ of the etalon. The latter together constitute the etalon's wavelength dependent optical path length $n\left(\lambda\right) \cdot L$. Here, $c_0$ is the speed of light in vacuum. We call the frequency interval corresponding to a phase change of $\Delta\varphi=2\pi$ the etalon's free spectral range (FSR), which is approximately 2\,GHz for the described setup. 
\par
In order to obtain an error signal for frequency stabilization of a laser, the interferometer signals are compared with electronically generated set signals $(U_{\mathrm{set},x}, U_{\mathrm{set},y}) = (\mathrm{cos} \,\varphi_{\mathrm{set}}, \mathrm{sin}\,\varphi_{\mathrm{set}})$ that depend on a single chosen set phase $\varphi_{\mathrm{set}}$. The difference between $\varphi$ and $\varphi_{\mathrm{set}}$ is used as a frequency error signal that varies with constant slope throughout the free spectral range. This allows to use the etalon for frequency stabilization of the laser at any point within the FSR. Further, tuning of a stabilized laser can be accomplished over a large wavelength range including many FSR by alteration of the generated $\varphi_{\mathrm{set}}$. 
\par
As the interferometer signals are periodic functions of the optical frequency, they are ambiguous. Hence, for stabilization to a specific target frequency one has to predetermine the laser's frequency to an accuracy of better than one half of an FSR. This can be achieved by means of a standard wavemeter. Further, $\varphi$ is subject to a drift of $n \cdot L$. We use a reference laser of precisely known frequency to keep track of and correct for small optical path length changes, that result in an offset of the measured phase at the given frequency. In practice, as typical interferometer drift rates are on the order of $100-200\,$kHz/min, this measurement has to be carried out about every $10\,$ minutes, which can be done automatically within approximately one to two seconds. With the help of these additional measures, each frequency is unambiguously assigned a phase, and continuous scans over an almost arbitrary amount of FSRs can be carried out.
\section{Calibration and characterization}
To reach an absolute frequency precision on the MHz level over a wavelength range of several tens of nanometers, three calibration steps are performed.
\par 
Firstly, a systematic deviation due to the approximation of the exact Airy function describing the interferometer signal as a sinusoidal function of the laser frequency in Eq.\,(\ref{phi_dependence}) has to be considered. This is in practice done by calibrating the set phase for a given target wavelength throughout one FSR. The calibration is achieved by comparing the frequency response of the interferometer with that of an etalon made of an optical fiber. With a fiber length of 2.5\,m the FSR of 60\,MHz is much smaller than the FSR of the interferometer. Therefore, the fiber etalon's fringe pattern varies much faster when the laser frequency is tuned as compared to the interferometer signals $U_x$ and $U_y$. This allows to use the fringes of the fiber etalon as equidistant markers for frequency intervals. In this way the deviations from the linear frequency dependence of the interferometer phase $\varphi$ can be measured. The maximum nonlinearity found for the quadrature interferometer is on the order of 2\% of one FSR, corresponding to a maximum of $40\,$MHz deviation in target frequencies within one etalon FSR of 2\,GHz. In order to compensate this systematic deviation in practice, the recorded data are electronically represented in a look-up-table (LUT). The LUT is used to modify the set phase $\varphi_{\mathrm{set}}$ for frequency tuning of the stabilized laser. With a set phase corrected by use of the LUT we find a remaining nonlinearity of 0.05\% using the same method. This corresponds to an accuracy of better than $1\,$MHz within one FSR. It is important to note that this procedure is easily repeated and an exact LUT can be created when the interferometer is used in a different wavelength range at a different refractive index of the etalon. However we found that one LUT remained valid when changing the frequency of the laser by several hundred FSR.
\begin{figure}[t]
\centerline{\includegraphics[width=\columnwidth]{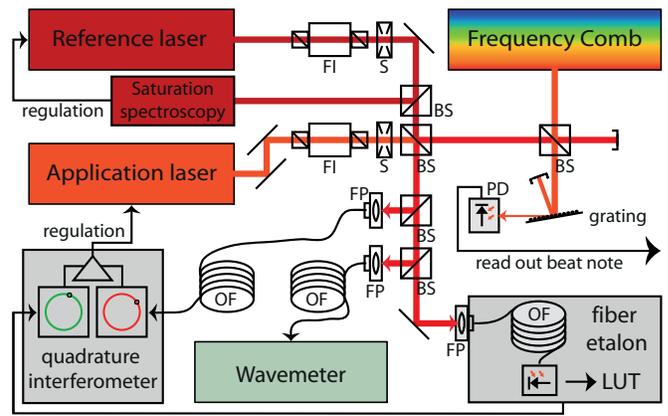}}
\caption{Setup for calibration and characterization of the interferometer. The frequency comb is a self referenced femto-second fiber laser based design (Menlo Systems). A Hydrogen maser and GPS are utilized as time standards. The fiber etalon is used to record the data used in the Look-Up-Table (LUT). A beat note with a frequency comb is recorded using a spectrum analyzer. Application laser: Toptica DLPro, Reference laser frequency $384.227981\,$THz (DFB diode laser, TEM Messtechnik), Saturation spectroscopy (Cosy, TEM Messtechnik), Wavemeter: High Finesse WS/7, FI: Faraday isolator, S: Shutter, BS: Beam Splitter, PD: PhotoDiode, OF: polarization maintaining Optical Fiber, FP: Fiber Port.}
\label{comb_measurement_setup}
\end{figure}
%
\par
\begin{figure}[b!]
\centerline{\includegraphics[width=\columnwidth]{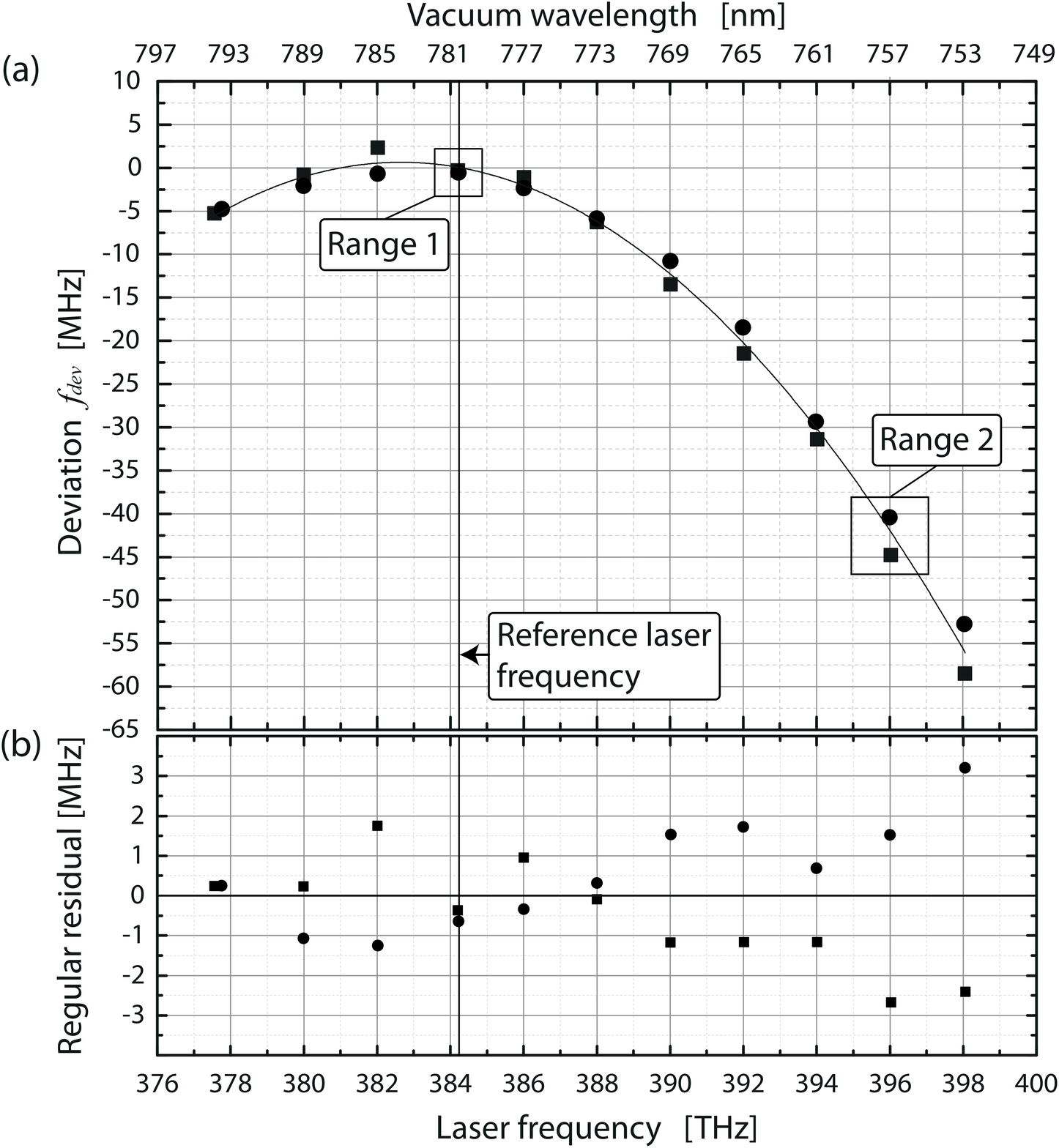}}
\caption{(a) measured frequency deviation $f_{dev}$ over the whole accessible wavelength range. Circles and squares are separate datasets taken on two consecutive days. One second order polynomial fit to both data sets is shown. The measurements for the frequency ranges 1 and 2 are shown enlarged in Fig.\,\ref{zoom-in}. (b) residual of a fit to both datasets shown in (a). A vertical line marks the reference laser frequency in both graphs.}
\label{wavelength_scan}
\end{figure}
\begin{figure}[t!]
\centerline{\includegraphics[width=\columnwidth]{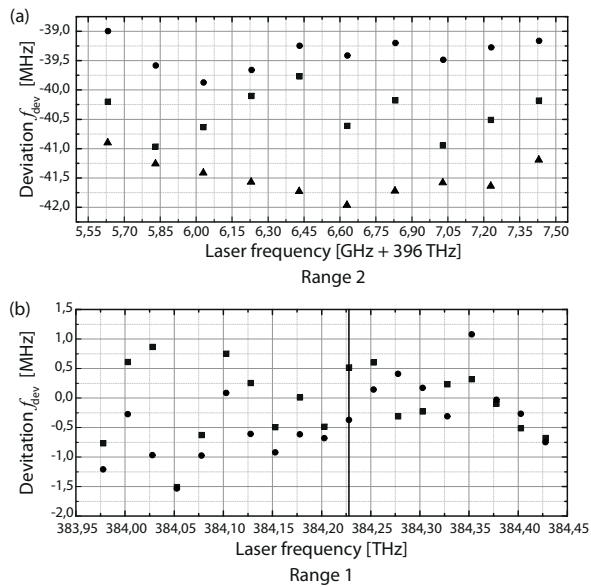}}
\caption{Frequency ranges 1 and 2 shown with higher resolution. (a): Frequency range 2. Three sets of data points at frequency $396\,$THz within one FSR of the etalon are shown. Circles and squares are taken using different offset phase measurements but the same  LUT. For the triangles a new LUT was generated and the offset phase was newly measured. A linear drift of $130\,$kHz/min obtained from repetitive measurements at the reference wavelength has been subtracted from all datasets for clearer visibility of the frequency-dependent error. (b): Frequency range 1. Two datasets close to the reference laser frequency (vertical line) are shown. To record this data, the reference laser (DFB diode laser) was controlled by the quadrature interferometer and interchanged with the application laser (Toptica DLPro).}
\label{zoom-in}
\end{figure}

In the following we describe a second calibration step that enables precise tuning over nanometers throughout a wavelength range as wide as $750 - 795\,$nm. Over such a broad wavelength range the dispersion of the etalon's medium is significant. The index of refraction can be described by the Sellmeier equation \cite{born_wolf_optics} with the specific material parameters for the BK7 etalon used in this case. However, in order to be able to calculate $\varphi_{\mathrm{set}}$ based on Eq.\,\ref{phi_dependence} with a precision corresponding to 1\,MHz, $n(\lambda)\cdot L$ has to be determined with a relative error on the order of $10^{-9}$. In this calibration step we obtain a precise measurement of the optical path $n(\lambda)\cdot L$ with the measurement setup shown in Fig.\,\ref{comb_measurement_setup}. For an absolute frequency reference we use a reference laser at the D2 optical transition line of $^{87}$Rb at $384.227981\,$THz \cite{springerlink:10.1007/BF00330229}. This laser is stabilized to the cross-over signal between the $5 ^2\rm{S}_{1/2},\, F=2 \rightarrow 5 ^2\rm{P}_{3/2},\, F=2,3$ transitions by means of a saturation spectroscopy. For a precise determination of the frequency of the application laser controlled by the interferometer setup we employ a frequency comb. The frequency is inferred by recording the beat note of the application laser with the comb, in combination with wavelength measurement carried out by a high resolution wavemeter. In order to collect data over a wide wavelength range the following steps have to be repeatedly performed: 
\begin{itemize}
\item The application laser is frequency stabilized to the interferometer with a given phase $\varphi_{\mathrm{set}}$ at a given wavelength.
\item The fiber etalon is used to create an LUT as explained in the first step.
\item The reference laser is used to perform a measurement of the interferometer phase offset. 
\item A precise measurement of the application laser frequency is carried out by means of the frequency comb for various $\varphi_{\mathrm{set}}$ around the given wavelength.
\end{itemize}
The results of this procedure for different wavelengths are summarized in Figs.\, \ref{wavelength_scan}(a) and \ref{zoom-in}. To obtain the remaining frequency deviation after applying the previous calibration step we process the data in the following way:
We first use Eq.\,\ref{phi_dependence} to convert the measured frequencies to interferometer phases and compare these to the set phases $\varphi_{\mathrm{set}}$. This resulting wavelength dependent phase deviation can be expressed as a frequency deviation $f_{\mathrm{dev}}$ by scaling with the etalon's FSR. The resulting data exhibit systematic deviations in the frequency control of the application laser on the order of several ten MHz over a wavelength range of 45\,nm. This corresponds to a relative error on the order of $10^{-7}$.
\par
In order to further characterize the performance of the device we show in Fig.\,\ref{zoom-in} the same data resolved for the two wavelength ranges that are indicated in Fig.\,\ref{wavelength_scan}\,(a):
\par
Fig.\,\ref{zoom-in}\,(a) shows the frequency deviation occurring within one FSR. This gives a measure on how well we are able to calibrate the relation between interferometer phase and frequency by means of the LUT. For a single scan we find small rms deviations. However, for repeated calibrations and subsequent scans over one FSR the precision of the device is limited by the reproducibility to an rms value of 0.9\,MHz. This can be attributed to frequency offsets consistent with the linewidth of the reference laser.
\par
Fig.\,\ref{zoom-in}\,(b) demonstrates the capability of precise frequency tuning over a range as broad as of $225$  FSRs. For this measurement a distributed feedback diode is used as an application laser, which was continuously stabilized throughout the scan. Again using the dataset shown as squares for calibration, we find an rms deviation of $0.7\,$MHz for the dataset represented with circles. Thus, the calibration step described above proofs valid for the complete wavelength range over approximately 1\,nm.
\par
To calculate the interferometer phase from the measured frequency in the above mentioned analysis we use the refractive index as described by the Sellmeier equation with coefficients specified for the BK7 in use, together with an estimated etalon length $L$. The optical path length has been estimated such that the observed deviations are minimal throughout the full wavelength range shown in Fig.\,\ref{wavelength_scan}\,(a). As a constraint the optical path length was chosen such that the phase offset measured at the wavelength of the reference laser was reproduced. With this the second calibration step, in which $n(\lambda)\cdot L$ is determined to a very high precision, is achieved.
\par
The characterization obtained with the data shown in Fig.\,\ref{wavelength_scan}\,(a) can be utilized for a third and last calibration step. For practical purposes we use a best fit to the data by a second order polynomial that serves as a calibration curve and is shown in Fig.\,\ref{wavelength_scan}\,(a). The residuals of this fit are shown in Fig.\,\ref{wavelength_scan}\,(b). Based on the obtained curve we can implement the calibration of the interferometer setup by applying a wavelength dependent correction for the refractive index up to second order. This effectively changes the etalon's optical path length used in our calculation of the interferometer phase for a target frequency. 
\par
To further demonstrate the effectiveness of this calibration we use a different second order polynomial fit function that has been obtained from the dataset represented as circles in Fig.\,\ref{wavelength_scan}\,(a). In this manner we find a maximum frequency deviation of $5.7\,$MHz (rms $1.6\,$MHz) in the dataset represented as squares for the fully calibrated interferometer. Comparison of two datasets obtained on consecutive days (circles, squares in Fig.\,\ref{wavelength_scan}) indicate a small systematic drift that is not further investigated within this work.
\section{Discussion}
It can be seen in the data presented in Fig.\,\ref{wavelength_scan}\,(a), that the frequency deviation is found to be dominantly of second order. This could occur due to a second order error in the wavelength dependence of the refractive index described with the specified Sellmeier parameters. In the second calibration step, we have chosen an optical path length which minimizes the overall frequency deviation in the whole wavelength range and leads to zero frequency deviation at the reference laser frequency. With this choice of $n \cdot L$, we compensate the zeroth order error in the description of the refractive index and obtain corrections in higher orders to the wavelength dependence of the refractive index. We find relative errors of  $4.4 \times 10^{-5} \pm 1.6 \times 10^{-6}$ in the first order and $8.0 \times 10^{-2} \pm 2.9 \times 10^{-3}$ in the second order. These deviations are consistent with a relative uncertainty of the refractive index described by the Sellmeier parameters of $2 \times 10^{-7}$.   
\par
As we have shown that a correction of the refractive index up to second order serves very well to calibrate the setup, it will from now on only be necessary to measure the deviation at three roughly equally spaced frequencies within the wavelength range. The calibration is then obtained by using a second order polynomial fit to these data points in the way as described above. In the examined wavelength interval, Doppler-free spectroscopies of optical transitions in rubidium and potassium are readily available and can serve as absolute frequency references. This eliminates the need for a frequency comb to carry out the last calibration step. In addition, employing data at just three wavelengths allows for efficient recalibration in order to compensate possible slow systematic drifts on the time scale of days. 
\section{Conclusion}
In this paper we describe a calibration procedure for an interferometric laser frequency stabilization in the wavelength range $750-795\,$nm. Several calibration steps are taken resulting in an exceptional precision with an rms deviation of $1.6\,$MHz. The maximum observed deviation of 5.7\,MHz corresponds to a relative accuracy of $1.4 \times 10^{-8}$. While the accuracy within one FSR is increased by approximately one order of magnitude by the first calibration step, a calibration over a wide frequency range allows the accurate determination of the interferometer's optical path length. Only this increases the precision of the device from hundreds of MHz to the MHz level over the wide frequency range. The calibration method presented here is expected to be applicable throughout the $250\,$nm wavelength range of the interferometer setup, which is limited by the choice of the single mode optical fibers used in the system. Further, continuous high bandwidth laser frequency stabilization over a wide tuning range is supported. This makes the presented laser frequency stabilization an optimal choice for the spectroscopy of molecules over a large wavelength range.
\par
The accuracy of the device is enhanced, if frequency stabilization in a comparatively smaller wavelength range of approximately 1\,nm is required, as shown in Fig.\,\ref{zoom-in}\,(b). 
If in this reference range a reference laser is available, the third calibration step can be omitted. If  for the respective wavelength range no laser spectroscopy is available as a reference, a beat note with a known frequency comb mode can be used as a reference. This allows then to continuously frequency stabilize the application laser while tuning over a range spanning hundreds of etalon FSRs of 2\,GHz. In comparison, a direct beat note of the application laser with the comb mode can be applied continuously only throughout a fraction of the narrow mode spacing of the comb.
\section*{Acknowledgments}
We thank Th. Udem and T. W. H\"ansch at the Max-Planck-Institute of Quantum Optics and Ronald Holzwarth at Menlo Systems GmbH for discussions and for making a frequency comb available.
%


\end{document}